\documentclass[aps,twocolumn]{revtex4}
\usepackage{graphicx}
\def\bea{\begin{eqnarray}}
\def\eea{\end{eqnarray}}
\def\nn{\nonumber}
\def\2{\hspace*{2mm}}
\def\3{\hspace*{3mm}}
\def\5{\hspace*{5mm}}
\def\h{\hspace*{0.5mm}}
\def\hh{\hspace*{0.25mm}}
\def\w{\vspace*{-2.1mm}}
\def\ww{\vspace*{-2.6mm}}
\def\br{{\bf r}}
\def\bk{{\bf k}}
\def\dr{{\rm d}{\bf r}}
\def\dk{{\rm d}{\bf k}}
\def\vv{|\hspace*{-.4mm}|}
\def\rr{\rangle\hspace*{-.75mm}\rangle}
\def\la{\langle}
\def\ra{\rangle}
\begin{document}

\title{Exact Symmetries of Electron Interactions in the Lowest Landau Level}

\author{M. Eliashvili}
\affiliation{Department of Theoretical Physics,
             A. Razmadze Mathematical Institute, Tbilisi 380093 Georgia}

\author{Z. F. Ezawa}
\affiliation{Department of Physics, Tohoku University, Sendai 980-8578 Japan}

\author{G. Tsitsishvili}
\affiliation{Department of Theoretical Physics,
             A$.$ Razmadze Mathematical Institute, Tbilisi 380093 Georgia}

\begin{abstract}\noindent
Considering the system of interacting electrons in the lowest Landau
level we show that the corresponding four-fermion Hamiltonian is
invariant with respect to the local area-preserving transformations.
Testing a certain class of interaction potentials, we find that
this symmetry is universal with respect to a concrete type of potentials.

\end{abstract}

\maketitle

\section{introduction}

The fractional quantum Hall effects \cite{ezawa,girvin} emerge
due to the strong electron-electron interactions.
In the corresponding theoretical constructions these interactions
give rise to certain difficulties because of their nonlinear
character. Therefore, the usual tool for studying the interaction
effects is an approximate analysis or the numerical calculus
(see e.g. \cite{fano} and references therein).

In the present paper we carry out an exact analysis of
electron-electron interactions in the lowest Landau level
(LLL) and reveal the exact symmetries of the corresponding
four-fermion Hamiltonian.

Symmetries strongly affect the properties of the physical constituents
such as the ground states, elementary excitations, etc. Therefore, the
matter of symmetries is an essential question, especially in the case
of nonlinear theories where the standard methods of studying
spectra and eigenstates become inefficient.

The main results of the paper are summarized in the following
two points:
\begin{itemize}
\item[(a)] The many-body Hamiltonian describing
{\it interacting electrons} in the LLL is invariant with respect to
the {\it local} area-preserving transformations for a ceratin class
of interaction potentials. This assertion holds for the systems with
a finite number of electrons as well as in the {\it thermodynamic limit}.
\item[(b)] The algebraic structure of this symmetry is insensitive
to a concrete type of potentials within the given class of interaction
potentials.
\end{itemize}
It is to be emphasized that the area-preserving transformations have
so far been known only to comprise the symmetries of many-body
Hamiltonian of {\it noninteracting electrons} in a homogeneous magnetic
field \cite{cappelli,karabali}.

The symmetry stated in point (a) leads to the degeneracy of the corresponding
energy levels and endows the eigenstates with specific algebraic structure.
For the purpose of verification and in order to outline these features in
a relatively transparent way we carry out the exact analytic diagonalization
for some few-body quantum states. The symmetry of electron-electron
interactions gives rise to the corresponding quantum number and subsequently,
to a novel classification of many-body quantum states in the LLL.

In section II we comment more on our main result together with its
physical implications. The detailed verification of the existence
of the symmetry is presented in section III. In order to shed more
light on the structure of spectrum and eigenstates of the system
under consideration, we perform an exact analytic diagonalization
in section IV for a finite number of interacting electrons.
In section V we present discussions.

\section{symmetry of interacting electrons}

Particles in the LLL carry the vanishing kinetic energy. Hence, the
Hamiltonian consists of a pure interaction of the type
\bea
H=\int\psi^*(\br)\h\psi^*(\br')\h V(|\hh\br-\br'|)\h
\psi(\br')\h\psi(\br)\h\dr\h\dr'
\eea
where $\psi(\br)$ is the electron field operator in the LLL, and
$V(r)$ is the interaction potential which is assumed to be
rotationally invariant. Below we consider some class of potentials
including the Coulomb interaction.

Adopting the circular geometry we present the electron field operator as
\bea
\psi(\br)=\sum_{n=0}^\infty c(n)\hh\psi_n(\br)
\eea
where
\bea
\psi_n({\bf r})=\h\frac{z^n\hh e^{-\hh\bar zz/2}}{\sqrt{2\pi n!}}
\eea
are the one-particle wave functions in the LLL. We employ
the magnetic length scale and use the dimensionless quantities
$\sqrt{2}\hh z=x+i\hh y$ and $\sqrt{2}\hh\bar z=x-i\hh y$.
Fermi amplitudes satisfy the standard anticommutation relations
$\{c(m)\hh,c^*(n)\}=\delta_{mn}$.

The algebra of area-preserving transformations usually referred to as
$W_\infty$ can be realized separately within each Landau level. The
corresponding LLL generators written in the second
quantized form appear as \cite{cappelli}
$$
{\cal L}_{mn}=\int\psi^*(\br)
\left[\frac{z}{2}-\frac{\partial}{\partial\bar z}\right]^m
\left[\frac{\bar z}{2}+\frac{\partial}{\partial z}\right]^n
\psi(\br)\h\dr,
$$
where $m$, $n$ are nonnegative integers, and $\psi(\br)$ is given by (2)
and (3).

Since ${\cal L}_{mn}$ involve the higher powers of derivatives, the
corresponding transformations are in general nonlocal. In contrast,
the elements
$W_L\equiv\{{\cal L}_{00},{\cal L}_{11},{\cal L}_{01},{\cal L}_{10}\}$
form the maximal subalgebra in $W_\infty$ generating the local
transformations. The main purpose of the present paper is to manifest
the invariance of the Hamiltonian (1) under these transformations, as
expressed by the commutation relations $[H,W_L]=0$.

The elements ${\cal L}_{00}$ and ${\cal L}_{11}$ represent the particle
number and total angular momentum operators, respectively
\bea
N\equiv{\cal L}_{00}=\sum_{n=0}^\infty c^*(n)\hh c(n),
\eea
\bea
M\equiv{\cal L}_{11}=\sum_{n=0}^\infty n\hh c^*(n)\hh c(n).
\eea
The relations $[H,{\cal L}_{00}]=0$ and $[H,{\cal L}_{11}]=0$
expressing the corresponding conservation laws are obvious.
The later takes place due to the rotational invariance of $V(r)$.
The rest part of $W_L$ is given by
$$
G^-\equiv{\cal L}_{10}=\sum_{n=0}^\infty\sqrt{n+1}\h\h c^*(n)\hh c(n+1),
$$
$$
G^+\equiv{\cal L}_{01}=\sum_{n=0}^\infty\sqrt{n+1}\h\h c^*(n+1)\hh c(n).
$$

In the succeeding section we prove the the relations
\bea
[H,G^\pm]=0
\eea
for the class of interaction potentials set by $V(r)=r^{-2\sigma}$
and $0<\sigma<3/4$. In the rest of this section we point
out some of their consequences. Let us first remark that
the operator $N$ commutes with $G^\pm$ and $H$. Hence, we work
within a given $N$-particle sector where $N$ can be treated
as an ordinary number.

As a result of this symmetry, the energy levels of the interacting
Hamiltonian (1) are infinitely degenerate. From the relations (6)
it follows that the eigenstates belonging to a given energy level
can be obtained one from another by repetitive applications of
operators $G^\pm$. The commutation relation $[\hh G^-,\hh G^+]=N$
indicates that $G^\pm$ form the ordinary oscillator algebra.
Therefore, the eigenstates of a given energy level can be classified
with respect to the amount of $G$-quanta. The corresponding number
operator is given by $N_G=N^{-1}G^+G^-$. Each energy level~possesses
its own $G$-vacuum defined by $G^-\vv0\rr=0$. Indeed, the operator
$G^-$ lowers the total angular momentum and its repetitive action
on any eigenstate will eventually give zero due to the Fermi
statistics. All other states within a given energy level can be
obtained from the corresponding $\vv0\rr$ by repetitive action
of the operator $G^+$ carrying no energy, no particle number and the
unit angular momentum: $[H,G^\pm]=0$, $[N,G^\pm]=0$ and
$[M,G^\pm]=\pm G^\pm$. Some of the details become more pictorial
in Section IV where we comment on the system with finite
number of interacting electrons.

The complete set of commuting operators employed in applications
to quantum Hall systems is usually taken as $\{H,N,M\}$. Once the
relation $[H,G^\pm]=0$ is found out, any combination of $G^\pm$
can be taken (instead of $M$) for the classification of
quantum states. In the preceding paragraph we have discussed the
operator $N_G$. However, it commutes with $M$, i.e. leads to the
equivalent classification scheme.

A novel alternative emerges from the set of commuting operators
given by $\{H,N,G^-\}$. Namely, the $N$-particle eigenstates of
$H$ related with a given energy level can be classified in terms
of the coherent states. They are given by
$\vv w\rr=\exp\hh(w\h G^+)\hh\vv0\rr$ where $w$ is a complex
number, and $\vv0\rr$ is the $G$-vacuum. The corresponding
eigenvalue equation reads as $G^-\hh\vv w\rr=w\hh N\hh\vv w\rr$.
Since the operator $G^+$ increases the total angular momentum,
the coherent states occupy the infinite area. Therefore, the
physically interesting case for the application of such a
classification scheme will be the thermodynamic limit.

\section{proof}

In this section we give a proof of the relation $[H,G^\pm]=0$.
Due to the identity $[H,G^+]^*=[G^-,H]$ it is sufficient to deal with
only one of the generators $G^\pm$. Rewriting $H$ in terms of $c^*$
and $c$ we get
$$
H=\sum_{lmnt}V_{lmnt}\h c^*(l)\h c^*(n)\h c(t)\h c(m),
$$
\bea
V_{lmnt}=\int\bar\psi_l(\br)\h\psi_m(\br)\h
V\h\bar\psi_n(\br')\h\psi_t(\br')\h\dr\h\dr'.
\eea
Employing $\{c(m)\hh,c^*(n)\}=\delta_{mn}$ and $V_{lmnt}=V_{ntlm}$ we
perform some trivial manipulations and arrive at
\bea
[\h H\h,\h G^+\h]
&=&\sum\sqrt{m+1}\h\h V_{l,m+1,n,t}
\h c^*(l)\hh c^*(n)\hh c(t)\hh c(m)-\nn\\\nn\\
&-&\sum\sqrt{l}\h\h V_{l-1,m,n,t}
\h c^*(l)\hh c^*(n)\hh c(t)\hh c(m)
\eea
Using (3) and (7) we get
\bea
\sqrt{m+1}\h\h V_{l,m+1,n,t}&=&
\int\dr\h\bar\psi_l(\br)
\bigg[\bigg(\frac{z}{2}-\frac{\partial}{\partial\bar z}\bigg)
\psi_m(\br)\bigg]\times\nn\\\nn\\
&\times&\int\dr'\h
V(|\hh\br-\br'|)\h\bar\psi_n(\br')\h\psi_t(\br'),\nn\\\nn\\
\sqrt{l}\h\h V_{l-1,m,n,t}&=&
\int\dr\h \bigg[\bigg(\frac{z}{2}+\frac{\partial}{\partial\bar z}\bigg)
\bar\psi_l(\br)\bigg]\psi_m(\br)\h\times\nn\\\nn\\
&\times&\int\dr'\h V(|\hh\br-\br'|)\h\bar\psi_n(\br')\h\psi_t(\br').\nn
\eea
Plugging these relations in (8) and integrating by parts one gets
$$
[\h H\h,\h G^+\h]=2\sum\{X_{lmnt}-S_{lmnt}\}
\h c^*(l)\hh c^*(n)\hh c(t)\hh c(m),
$$
$$
X_{lmnt}=\int\bar\psi_l(\br)\h\psi_m(\br)\h
\frac{\partial V}{\partial\bar z}\h\h\bar\psi_n(\br')
\h\psi_t(\br')\h\dr\h\dr',
$$
$$
S_{lmnt}=\int\frac{\partial}{\partial\bar z}\h\big[\h\bar\psi_l(\br)
\h\psi_m(\br)\h V\h\big]\h\bar\psi_n(\br')
\h\psi_t(\br')\h\dr\h\dr'.
$$
Interchanging the integration variables in $X_{lmnt}$ we use
$$
\frac{\partial}{\partial\bar z}\h V(|\hh\br-\br'|)=
-\frac{\partial}{\partial\bar z'}\h V(|\hh\br'-\br\hh|)
$$
and get $X_{lmnt}=-\h X_{ntlm}$. This forces the corresponding term in
$[H,G^+]$ to vanish and yields
$$
[\h H\h,\h G^+\h]=-2\sum S_{lmnt}\h c^*(l)\hh c^*(n)\hh c(t)\hh c(m).
$$

The integrand in $S_{lmnt}$ represents the total derivative with
respect to $\br$. Restricting the corresponding integration to a
finite disc of radius $R$, we subsequently pass to the surface
integral and perform the limit $R\to\infty$. This leads to
\begin{eqnarray}
\sqrt{2}\h S_{lmnt}&=&\oint d\theta\hh R\h e^{i\theta}\hh
\bar\psi_l({\bf R})\h\psi_m({\bf R})\times\nn\\\nn\\
&\times&
\int V(|{\bf R}-\br'|)\h\bar\psi_n(\br')
\h\psi_t(\br')\h\dr'
\end{eqnarray}
where ${\bf R}=\{R\cos\theta,R\sin\theta\hh\}$ with $R\to\infty$.

For the sake of convenience we use the Fourrier representation of
the interaction potential
\bea
V(r)=\frac{1}{2\pi}\int V(k)\h e^{i\hh\bk\br}\h\dk
\eea
where $\bk=\{k\cos\gamma,k\sin\gamma\}$. Besides, we assume
that $V(k)$ possesses the necessary behaviour at $k=0$ and
$k=\infty$, so that the Fourrier integral (10) is convergent.

Substituting (10) into (9) and performing all the necessary
integrations we get
\begin{eqnarray}
S_{lmnt}&=&\delta_{l+n,m+t+1}\frac{R^{l+m+1}\hh
e^{-R^2/2}}{\sqrt{l!\hh m!\hh2^{l+m+1}}}\times\nn\\\nn\\
&\times&\int_0^\infty J_{|l-m-1|}(kR)\hh\omega_{nt}(k)\hh
V(k)\hh k\hh dk\nn
\end{eqnarray}
Deriving the last expression we have used
$$
\oint e^{i\alpha\theta+i\hh kR\cos\theta}d\theta
=2\pi\hh i^{|\alpha|}J_{|\alpha|}(kR)
$$
\bea
\int e^{-i\hh\bk\br'}\hh\bar\psi_n(\br')
\h\psi_t(\br')\h\dr'=(-i)^{|n-t|}e^{-i\hh(n-t)\gamma}\omega_{nt}
\eea
where $J_n(kR)$ is the Bessel function, and the quantity
$\omega_{nt}=\omega_{tn}$ is given by
\bea
\omega_{n,n+\alpha}
=\frac{\sqrt{n!}}{\sqrt{(n+\alpha)!}}\h\h\bigg[\frac{k^2}{2}\bigg]^{\alpha/2}
e^{-k^2/2}\h L_n^\alpha(k^2/2)
\eea
with $L_n^\alpha$ denoting the Laguerre polynomials.

We consider $n\leq t=n+\alpha$ (with $\alpha=0,1,\ldots$)
separately from $t<n=t+\alpha$ (with $\alpha=1,2\ldots$) and get
\bea
S_{m+\alpha+1,m,n,n+\alpha}=\frac{L^{m+1+\alpha/2}\h e^{-L}}
{\sqrt{2^\alpha\hh m!\h(m+\alpha+1)!}}\h\h I_{n\alpha}(L)
\eea
\bea
S_{l,l+\alpha-1,t+\alpha,t}=\frac{L^{l+\alpha/2}\h e^{-L}}
{\sqrt{2^\alpha\hh l!\h(l+\alpha-1)!}}\h\h I_{t\alpha}(L)
\eea
\bea
I_{n\alpha}=\int_0^\infty V(k)J_\alpha(k\sqrt{2L})\h
k^{\alpha+1}
L^\alpha_n(k^2/2)\h e^{-k^2/2}\h dk\nn
\eea
where $L\equiv R^2/2$. In fact, $L$ sets the order of the maximal
one-particle angular momentum which can be accommodated within a finite
disk of radius $R$.

The two cases presented in (13) and (14) can be related with each
other as
\bea
\frac{S_{l,l+\alpha-1,t+\alpha,t}}{S_{l+\alpha+1,l,t,t+\alpha}}
=\frac{\sqrt{(l+\alpha)(l+\alpha+1)}}{L}
\eea
and therefore, the limit $L\to\infty$ may be studied for only
one of them, say for (13).

If the system contains finite number of electrons, then the quantities
$m$, $n$, $\alpha$ in (13) may take the finite values. This corresponds
to the finite total angular momentum. In that case the limiting
procedure $L\to\infty$ is trivial and due to the exponential factor
$e^{-L}$ leads to $S_{lmnt}=0$. On the other hand, even if the electron
number is finite, the total angular momentum may become infinite, and
$m$, $n$, $\alpha$ will take the infinite values. Physically most
interesting case where $m$, $n$, $\alpha$ necessarily take the infinite
values is the {\it thermodynamic limit}, i.e. when the electron number
increases together with $L$ forming the finite density state.
Thus, the limiting procedure should be considered in all possible cases
where any of $m$, $n$, $\alpha$ in (13) may become infinite together with $L$.
In that case the factor $e^{-L}$ may be compensated by other ones, and the
problem requires the detailed analysis.

The comprehensive analysis for general $V(r)$ is quite a nontrivial problem
and therefore, we concentrate on the particular type of potential given by
\bea
V(k)=\frac{\Gamma(1-\sigma)}{2^{2\sigma-1}\Gamma(\sigma)}
\h\frac{1}{k^{2-2\sigma}}
\eea
where $0<\sigma<3/4$. This restriction on $\sigma$ guarantees the integral
(10) to be convergent, and gives $V(r)=r^{-2\sigma}$. The Coulomb interaction
corresponds to $\sigma=1/2$.

In this case the integral in the r.h.s. of (13) leads to
(see 2.19.12.7 in Ref.[4])
\begin{widetext}
$$
S=\frac{2^{-\sigma}\hh\Gamma(n+1-\sigma)\hh\Gamma(\alpha+\sigma)\hh
L^{m+\alpha+1}\hh e^{-L}}{\Gamma(\sigma)\hh\Gamma(\alpha+1)
\sqrt{\h m!\h(m+\alpha+1)!\h n!\h(n+\alpha)!}}\h\h
_2F_2(\alpha+\sigma,\sigma;\sigma-n,\alpha+1;-L)
$$
where $S$ stands for $S_{m+\alpha+1,m,n,n+\alpha}$ and $_2F_2$ is the
hypergeometric function.

Using the integral representation, we express $_2F_2$ in terms
of $_1F_1$. Subsequently we perform the Kummer transformation
(13.1.27 in Ref.[5]) and arrive at
$$
S=\frac{2^{-\sigma}\hh\Gamma(n+1-\sigma)\hh L^{m+\alpha+1}\hh e^{-L}}
{\Gamma(\sigma)\hh\Gamma(1-\sigma)\hh
\sqrt{m!\h(m+\alpha+1)!\h n!\h(n+\alpha)!\h}}\h
\int_0^1
\tau^{\alpha+\sigma-1}(1-\tau)^{-\sigma}\h_1F_1(-n;\sigma-n;L\tau)\h
e^{-L\tau}\hh d\tau
$$
where $_1F_1(-n;\sigma-n;L\tau)$ is in fact the $n$'th order polynomial
with respect to $L\tau$. Using its power expansion we get
$$
S=\frac{2^{-\sigma}\hh L^{m+\alpha+1}\hh e^{-L}}
{\Gamma(\sigma)\hh\Gamma(1-\sigma)\hh
\sqrt{m!\h(m+\alpha+1)!\h n!\h(n+\alpha)!\h}}\h
\sum_{k=0}^nC_n^{\hh k}\h\Gamma(n-k+1-\sigma)\h L^k
\int_0^1\frac{\tau^{k+\alpha+\sigma-1}e^{-L\tau}}{(1-\tau)^\sigma}\h\h d\tau
$$
where $C_n^{\hh k}$ is the Newton binomial.

Consider now the integral in the last expression, and remark that
the quantity $k+\alpha$ may take infinite as well as finite values.
Hence, we need its ($L\to\infty$)-form valid for all (including
infinite) values of $k+\alpha$.

The factor $e^{-L\tau}$ provides the exponential suppression of all
contributions coming from $\tau\ne0$. The integrable pole
at $\tau=1$ also fails to contribute because of the same exponential
suppression. So, the integral gets the contribution from the vicinity
of $\tau=0$, and we may replace the factor $(1-\tau)^\sigma$ by its
value in $\tau=0$. In this consideration we get
$$
\int_0^1\frac{\tau^{k+\alpha+\sigma-1}e^{-L\tau}}{(1-\tau)^\sigma}\h\h
d\tau\to\int_0^1\tau^{k+\alpha+\sigma-1}e^{-L\tau}\h d\tau=
\frac{1}{L^{k+\alpha+\sigma}}\int_0^L\tau^{k+\alpha+\sigma-1}e^{-\tau}\h d\tau.
$$
If $k+\alpha+\sigma$ is finite, then for $L\to\infty$ we get
\bea
\int_0^1\frac{\tau^{k+\alpha+\sigma-1}e^{-L\tau}}{(1-\tau)^\sigma}\h\h d\tau
\2\to\2\frac{\Gamma(k+\alpha+\sigma)}{L^{k+\alpha+\sigma}}
\eea
This relation is valid also for $k+\alpha+\sigma\to\infty\h$. Indeed, we have
$$
\int_0^L\tau^{k+\alpha+\sigma-1}e^{-\tau}\h d\tau=
\int_0^1\tau^{k+\alpha+\sigma-1}e^{-\tau}\h d\tau+
\int_1^L\tau^{k+\alpha+\sigma-1}e^{-\tau}\h d\tau
$$
Remark, that both of the limits $L$ and $k+\alpha+\sigma$ in the second term
bring the divergent effects. Therefore, one may account them in any desired
order, say first $L\to\infty$ and then $k+\alpha+\sigma\to\infty$. This leads
back to (17), and hence to
$$
S=\frac{[\hh2^\sigma\hh\Gamma(\sigma)\hh\Gamma(1-\sigma)\hh]^{-1}L^{m+1-\sigma}\hh e^{-L}}
{\sqrt{m!\h(m+\alpha+1)!\h n!\h(n+\alpha)!\h}}\h
\sum_{k=0}^nC_n^{\hh k}\h\Gamma(n-k+1-\sigma)\h\Gamma(k+\alpha+\sigma).
$$
Using
$$
\Gamma(a)\Gamma(b)=\Gamma(a+b)\int_0^1\tau^{a-1}(1-\tau)^{b-1}d\tau
$$
the summation over $k$ can be performed exactly and for $S$ we get
\bea
S_{m+\alpha+1,m,n,n+\alpha}=\frac{2^{-\sigma}}{\Gamma(\sigma)}
\frac{L^{m+1-\sigma}e^{-L}}{\sqrt{m!\h(m+\alpha+1)!}}\h
\sqrt\frac{(n+\alpha)!}{n!}\h
\frac{\Gamma(\alpha+\sigma)}{\Gamma(\alpha+1)}
\eea
\end{widetext}
where any kind of limiting procedure is straightforward.

As an example we present the case when all of $m$, $n$,
$\alpha$ increase together with $L$. Introduce the parametrization
\bea
m=\lambda_mL\5\5n=\lambda_nL\5\5\alpha=\lambda_\alpha L
\eea
where $\lambda_{m,n,\alpha}$ are positive constants (none of them
may vanish, otherwise the corresponding momentum will stay finite
instead of tending to infinity). Further, the subscripts in
$S_{m+\alpha+1,m,n,n+\alpha}$ represent the one-particle angular
momenta, whose maximal values are of order of $L$, while the matter
is confined to the interior (including the edge) of the disk. Hence,
$\lambda_{m,n,\alpha}$ must be restricted as
\bea
0<\lambda_m+\lambda_\alpha\leq1
\eea
\bea
0<\lambda_n+\lambda_\alpha\leq1
\eea
\bea
0<\lambda_{m,n,\alpha}<1
\eea
Remark, that in contrast with $\lambda_m+\lambda_\alpha$ and
$\lambda_n+\lambda_\alpha$, the constants $\lambda_{m,n,\alpha}$
cannot be equal to one, since this would contradict to (20)
and (21) due to $\lambda_{m,n,\alpha}>0$.

We substitute (19) into (18) and employing Stirling's formula
arrive at
\bea
S_{m+\alpha+1,m,n,n+\alpha}=CL^{-1}\h e^{-LW}\nn
\eea
where $C$ is an $L$-independent constant, and $W$ looks as
$$
W(\lambda_m,\lambda_n,\lambda_\alpha)\h=\h1-\lambda_m
+\frac12\h\lambda_m\ln\lambda_m+\frac12\h\lambda_n\ln\lambda_n+\nn
$$
$$
+\frac12\h(\lambda_m+\lambda_\alpha)\ln(\lambda_m+\lambda_\alpha)
-\frac12\h(\lambda_n+\lambda_\alpha)\ln(\lambda_n+\lambda_\alpha).\nn
$$

The final step is to show that $W>0$. Remark,  that
$$
\frac{\partial W}{\partial\lambda_m}=
\frac12\ln[\lambda_m(\lambda_m+\lambda_\alpha)]<0,
$$
$$
\frac{\partial W}{\partial\lambda_n}=
\frac12\ln\frac{\lambda_n}{\lambda_n+\lambda_\alpha}<0,
$$
$$
\lambda_m\leq1-\lambda_\alpha\hh,\5\5
\lambda_n\leq1-\lambda_\alpha,
$$
where the last two relations follow from (20) and (21). Consequently,
$W(\lambda_m,\lambda_n,\lambda_\alpha)\geq
W(1-\lambda_\alpha,1-\lambda_\alpha,\lambda_\alpha)$ where
$$
W(1-\lambda_\alpha,1-\lambda_\alpha,\lambda_\alpha)=
\lambda_\alpha+(1-\lambda_\alpha)\ln(1-\lambda_\alpha).
$$
For $0<z<1$ we have $z\ln z>z-1$. Consequently,
$W(1-\lambda_\alpha,1-\lambda_\alpha,\lambda_\alpha)>0$
and $W(\lambda_m,\lambda_n,\lambda_\alpha)>0$. Hence,
$S_{m+\alpha+1,m,n,n+\alpha}=0$, and further, due to (15)
$S_{lmnt}=0$. In the similar way we show that $S_{lmnt}$ vanishes in all
possible cases when some (or none) of $m$, $n$, $\alpha$ tend to infinity.
Thus, for the interacting electron system defined by (1) -- (3) and (16)
we have demonstrated the relation (6) i.e. the existence of symmetries
with respect to the local area-preserving transformation.

\section{diagonalization}

Due to the symmetry established in the previous sections, the eigenvalue
spectrum of the model is highly degenerated, while the eigenstates exhibit
a certain kind of structure. In order to outline these features in a
relatively transparent way, we present an exact analytic diagonalization
procedure.

Since the interaction preserves the number of particles, the procedure
can be carried out in a different $N$-particle Fock subspaces independently.
The basis is chosen to be
$|\hh k_1\dots\hh k_N\ra\equiv c^*(k_1)\cdots\hh c^*(k_N)\hh|\hh0\hh\ra$
where $k_1<\cdots<k_N$.

Treating $\{\hh k_1\ldots\hh k_N\}$ as a multi-index we consider the quantity
$\la k_1\ldots k_N|H|\hh l_1\ldots l_N\ra$ as an $\infty\times\infty$
square matrix to be diagonalized.
Due to the circular symmetry of $V(r)$ we have
$$
\la k_1\dots\hh k_N| H\hh |\hh l_1\dots\hh l_N\ra\propto
\delta_{k_1+\cdots+k_N,l_1+\cdots+l_N}
$$
reflecting the angular momentum conservation. Hence, the matrix
$\la k_1\ldots k_N|H|\hh l_1\ldots l_N\ra$ has the block-diagonal
form, where each separate block is a finite-dimensional square
matrix labelled with some fixed value of the total angular momentum
$M=k_1+\cdots+k_N=l_1+\cdots+l_N$. We regard the blocks to be
arranged in the increasing order with respect to $M$, starting
from $M_{min}=N(N-1)/2$. The typical structure is shown in
Figure 1 exhibiting the lowest momentum blocks in
$\la k_1k_2\hh k_3|H|\hh l_1l_2\hh l_3\ra$.

\begin{figure}
\includegraphics{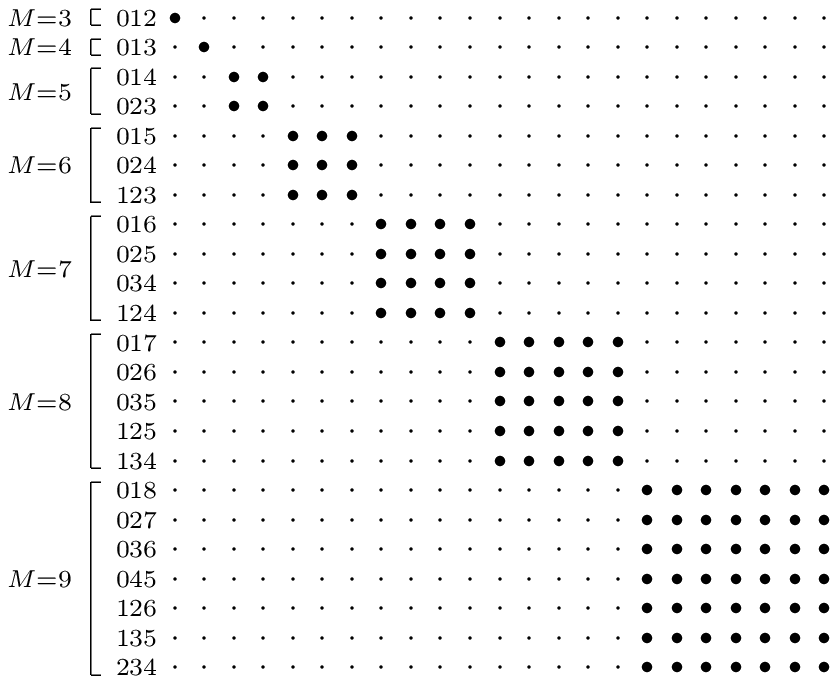}
\caption{Low-momentum structures of
$\la k_1k_2\hh k_3|H|\hh l_1l_2\hh l_3\ra$.
Bullets ($\bullet$) denote the nonvanishing matrix
elements, while the dots ($\cdot$) stand for zeroes.}
\end{figure}

Denote the block dimensionality by $d(N,M)$. Proceeding from the Fermi
statistics, the quantity $d(N,M)$ represents the number of partitions
of the integer $M$ into $N$ distinct nonnegative integers.
It is given via the following formal expansion \cite{andrews}
\bea
\sum_{M=0}^\infty d(N,M)\h
\tau^M=\frac{\tau^{N(N-1)/2}}{(1-\tau)(1-\tau^2)\cdots(1-\tau^N)}
\eea
where the factor $\tau^{N(N-1)/2}$ is related with the minimal
value of $M$.

Thus, the general $(N,M)$-block comprises the $d(N,M)$ number of
eigenvalues and eigenvectors. Denoting them by $E_{NM\beta}$ and
$\theta_{M\beta}(k_1\dots\h k_N)$ with $\beta=1,\ldots,d$ we
construct the Hamiltonian eigenstates as
$$
\vv N,M,\beta\hh\rr=\sum_{k_1<}\cdots\sum_{<k_N}
\theta_{M\beta}(k_1\dots\h k_N)
\hh\big|\hh k_1\dots\h k_N\big\ra.
$$

Each block carries some fixed value of the filling factor defined by
$\nu=N/K$ where $K$ is the number of available one-particle states
in a block. It is given by $K=1+l_{max}$ where $l_{max}$ is the
maximal value of one-particle angular momentum in a block. The later
can be calculated from
$$
\underbrace{0}_{1'st}+\underbrace{1}_{2'nd}+\cdots+
\underbrace{N-2}_{(N-1)'th}+\underbrace{l_{max}}_{N'th}=M.
$$
Hence, the filling factor for a given block appears as
$$
\nu=\frac{N}{M+1-(N-1)(N-2)/2}=\frac{1}{1+{\cal M}/N}
$$
where ${\cal M}=M-M_{min}$ is the relative momentum.

We move on to the results. Substituting (10) into (7) and using
(11) one arrives at
$$
V_{mlnt}=\delta_{m+n,l+t}\int_0^\infty
V(k)\h\omega_{ml}(k)\h\omega_{nt}(k)\h k\h dk,
$$
where $\omega_{ml}$ is given by (12). We use (16) and replace
the Laguerre polynomials in $\omega_{ml}$ and $\omega_{nt}$
by the corresponding power expansions. The integrals become
trivial, and $V_{mlnt}$ takes the form of a double finite sum.
The later is easily calculable if the one-particle angular
momenta $m$, $n$, $l$, $t$ are confined to some reasonable range,
say as in the blocks presented in Figure 1. For $V(r)$ given
by (16) (with an appropriate normalization) the matrix elements
take the rational (exact) values. Hence, the corresponding secular
equations can be written in the exact form, as well. In such a way
we obtain the exact eigenvalues.

In the Table I we present the eigenvalues of the blocks with
$M=3\div8$ shown in Figure 1. The case with $M=9$ is shown
separately.  The eigenvalues are labelled with index $\beta$ in
the order of their appearance when increasing the quantity $M$.

\begin{table*}
\caption{The values of some $E_{N=3,M\beta}$.
Superscripts in square braces stand for the values of $\beta$.
The constants $1/A=3\Gamma(1/2)$ and $1/B=2^{1/3}\Gamma(2/3)$
are introduced for the purpose of convenience.
$E^{[6,7]}$ are given by (24) and (25) in the text.}
\begin{ruledtabular}
\begin{tabular}{l|ccccccc|ccccccc}
&&&$V(r)\hspace*{-15.5mm}$&$=\hspace*{-7mm}$&$\hspace*{-6mm}\displaystyle{\frac{A}{r}}$&&
&&&$V(r)\hspace*{-14mm}$&$=\hspace*{-4mm}$&$\hspace*{-8mm}\displaystyle{\frac{B}{r^{1/3}}}$&&\\
\vspace*{-4mm}&&&&&&&&&&&&&&\\
\hline
\vspace*{-4mm}&&&&&&&&&&&&&&\\
\w
&&&&&&&&&&&&&&\\
$M=3\3$&$\3\frac{29}{32}^{[1]}$&&&&&&&
$\3\frac{101}{54}^{[1]}$&&&&&&\\
\w&&&&&&&&&&&&&&\\
$M=4\3$&$\3\frac{29}{32}^{[1]}$&&&&&&&
$\3\frac{101}{54}^{[1]}$&&&&&&\\
\w&&&&&&&&&&&&&&\\
$M=5\3$&$\3\frac{29}{32}^{[1]}$&$\frac{1627}{2048}^{[2]}$&&&&&&
$\3\frac{101}{54}^{[1]}$&$\frac{3317}{1944}^{[2]}$&&&&&\\
\w&&&&&&&&&&&&&&\\
$M=6\3$&$\3\frac{29}{32}^{[1]}$&$\frac{1627}{2048}^{[2]}$&$\frac{1373}{2048}^{[3]}$&&&&&
$\3\frac{101}{54}^{[1]}$&$\frac{3317}{1944}^{[2]}$&$\frac{997}{648}^{[3]}$&&&&\\
\w&&&&&&&&&&&&&&\\
$M=7\3$&$\3\frac{29}{32}^{[1]}$&$\frac{1627}{2048}^{[2]}$&
$\frac{1373}{2048}^{[3]}$&$\frac{46761}{65536}^{[4]}$&&&&
$\3\frac{101}{54}^{[1]}$&$\frac{3317}{1944}^{[2]}$&
$\frac{997}{648}^{[3]}$&$\frac{110915}{69984}^{[4]}$&&&\\
\w&&&&&&&&&&&&&&\\
$M=8\3$&$\3\frac{29}{32}^{[1]}$&$\frac{1627}{2048}^{[2]}$&
$\frac{1373}{2048}^{[3]}$&$\frac{46761}{65536}^{[4]}$& $\frac{21085}{32768}^{[5]}$&&&
$\3\frac{101}{54}^{[1]}$&$\frac{3317}{1944}^{[2]}$&
$\frac{997}{648}^{[3]}$&$\frac{110915}{69984}^{[4]}$&
$\frac{17315}{11664}^{[5]}$&&\\
\w&&&&&&&&&&&&&&\\
$M=9\3$&$\3\frac{29}{32}^{[1]}$&$\frac{1627}{2048}^{[2]}$&
$\frac{1373}{2048}^{[3]}$&$\frac{46761}{65536}^{[4]}$&
$\frac{21085}{32768}^{[5]}$&$E^{[6]}$&$E^{[7]}$\3&
$\3\frac{101}{54}^{[1]}$&$\frac{3317}{1944}^{[2]}$&
$\frac{997}{648}^{[3]}$&$\frac{110915}{69984}^{[4]}$&
$\frac{17315}{11664}^{[5]}$&$E^{[6]}$&$E^{[7]}$
\vspace*{1.5mm}
%\w
%&&&&&&&&&&&&&&
\end{tabular}
\end{ruledtabular}
\end{table*}

We observe that the spectrum of the block with total angular
momentum $M+1$ entirely comprises the one corresponding to $M$,
and also some additional eigenvalues. The amount of these
additional eigenvalues is given by $d(N,M+1)-d(N,M)$ and increases
together with $M$. As an example, passing from $M=8$ to $M=9$ in
$N=3$ we get two additional eigenvalues $E^{[6,7]}$ (see Table I
and Figure 1) given by
\bea
\frac{10025047\pm3\sqrt{107141413705}}{16777216}\h,\5
V(r)=\frac{A}{r}
\eea
\bea
\frac{96437129\pm\sqrt{27563293275793}}{68024448}\h,\5
V(r)=\frac{B}{r^{2/3}}
\eea
where the constants $A$ and $B$ are given in the caption of Table I.

Thus, once an eigenvalue appears in any block, it persists in all
subsequent ones and gives rise to an infinitely degenerated energy
level. The corresponding eigenstates are related with each other by
$$
G^\pm\vv N,M,\beta\hh\rr=\lambda^\pm(N,M,\beta)\hh\vv N,M\pm1,\beta\hh\rr
$$
where $\lambda^\pm$ are some constants.

Using these relations we can construct some of eigenstates without
solving out them explicitly. Consider the case of $N=3$ (see Figure 1).
The lowest block possesses one single eigenstate $\vv3,3,1\rr=|012\ra$.
The next block is also one-dimensional. The corresponding
eigenstate can be obtained by applying $G^+$ to $\vv3,3,1\rr$.
Since $G^+$ does not preserve the norm, its action should be
accompanied with the subsequent normalization. In such a way
we get $\vv3,4,1\rr=|013\ra$. The next block ($M=5$) possesses
two eigenstates. One of them can be obtained from $\vv3,4,1\rr$
by the action of $G^+$. It is given by
$$
\vv3,5,1\rr=\frac{\sqrt2}{\sqrt3}\h\h|014\ra
+\frac{1}{\sqrt3}\h\h|023\ra.
$$
The other one is fixed by the orthogonality principle,
and appears as
$$
\vv3,5,2\rr=\frac{1}{\sqrt3}\h\h|014\ra
-\frac{\sqrt2}{\sqrt3}\h\h|023\ra.
$$
In the similar way we can construct the eigenstates corresponding to
$M=6,7,8$.

The starting eigenstate $\vv3,3,1\rr\equiv|012\ra$, the operator
$G^+$ and the principle of orthogonality have nothing to do with
$V(r)$. Consequently, the eigenstates corresponding to $M=3\div8$
which can be constructed employing only these tools, do not depend
on the concrete type of $V(r)$. Passing to $M=9$ we acquire two
additional eigenstates (instead of a single one as for preceding blocks).
The principle of orthogonality becomes insufficient for fixing both
of them and leaves the one-parameter arbitrariness with respect
to rotation in the subspace of those two eigenstates. This degree
of freedom is governed by $V(r)$.

As we have remarked, the amount of additional eigenstates given by
$d(N,M+1)-d(N,M)$ increases together with $M$. Hence, the arbitrariness
left after the orthogonalization also increases, and the interaction
effects become more and more decisive in the formation of quantum states.

The only exception is the two-particle sector ($N=2$). Indeed, from (23)
we get
\bea
\sum_{M=0}^\infty d(2,M)\h
\tau^M&=&\frac{\tau}{(1-\tau)(1-\tau^2)}=
\frac{\tau(1+\tau)}{(1-\tau^2)^2}=\nn\\\nn\\
&=&\tau(1+\tau)(1+2\tau^2+3\tau^4+\cdots\h\h)\nn
\eea
indicating that not more than one additional eigenstate appears when
passing from $M$ to $M+1$. Thus, starting with $\vv2,1,1\rr=|01\ra$,
the operator $\hat G^+$ and the orthogonalization procedure completely
fix the set of two-particle eigenstates. So, they do not depend on the
concrete type of $V(r)$, i.e. are universal. Introducing the holomorphic
part $P(z_1,z_2)$ as
$$
\la\hh0\hh|\psi(\br_1)\psi(\br_2)\vv\hh2,M,\beta\hh\rr
=\h\frac{P(z_1,z_2)}{2\pi}\h\h e^{-\hh(1/2)(|z_1|^2+|z_2|^2)}
$$
we show some of two-particle eigenstates in Table II.

\begin{table}
\caption{Two-particle eigenstates with $M=1\div6$.}
\begin{ruledtabular}
\begin{tabular}{ll|ll}
\vspace*{-3mm}&&&\\
Fock representation&&&Holomorphic part\\
\ww&&&\\
\hline
\ww&&&\\
$\vv2,1,1\rr=|01\ra$&&&$P=(z_1-z_2)$\\
\ww&&&\\
\hline
\ww&&&\\
$\vv2,2,1\rr=|02\ra$&&&$P=\frac{(z_1-z_2)(z_1+z_2)}{\sqrt2}$\\
\ww&&&\\
\hline
\ww&&&\\
$\vv2,3,1\rr=\frac{|03\ra+\sqrt3\h|12\ra}{2}$&&&$P=\frac{(z_1-z_2)(z_1+z_2)^2}{2\sqrt2}$\\
\ww&&&\\
$\vv2,3,2\rr=\frac{|03\ra-\sqrt3\h|12\ra}{2}$&&&$P=\frac{(z_1-z_2)^3}{2\sqrt2}$\\
\ww&&&\\
\hline
\ww&&&\\
$\vv2,4,1\rr=\frac{|04\ra+|13\ra}{\sqrt2}$&&&$P=\frac{(z_1-z_2)(z_1+z_2)^3}{4\sqrt3}$\\
\ww&&&\\
$\vv2,4,2\rr=\frac{|04\ra-|13\ra}{\sqrt2}$&&&$P=\frac{(z_1-z_2)^3(z_1+z_2)}{4\sqrt3}$\\
\ww&&&\\
\hline
\ww&&&\\
$\vv2,5,1\rr=\frac{\sqrt5\h|05\ra+3\h|14\ra+\sqrt2\h|23\ra}{4}$&&&$P=\frac{(z_1-z_2)(z_1+z_2)^4}{8\sqrt6}$\\
\ww&&&\\
$\vv2,5,2\rr=\frac{\sqrt5\h|05\ra-|14\ra-\sqrt2\h|23\ra}{2\sqrt2}$&&&$P=\frac{(z_1-z_2)^3(z_1+z_2)^2}{8\sqrt3}$\\
\ww&&&\\
$\vv2,5,3\rr=\frac{|05\ra-\sqrt5\h|14\ra+\sqrt{10}\h|23\ra}{4}$&&&$P=\frac{(z_1-z_2)^5}{8\sqrt{30}}$\\
\ww&&&\\
\hline
\ww&&&\\
$\vv2,6,1\rr=\frac{\sqrt3\h|06\ra+2\sqrt2\h|15\ra+\sqrt5\h|24\ra}{4}$&&&$P=\frac{(z_1-z_2)(z_1+z_2)^5}{16\sqrt{15}}$\\
\ww&&&\\
$\vv2,6,2\rr=\frac{\sqrt5\h|06\ra+0\cdot|15\ra-\sqrt3\h|24\ra}{2\sqrt2}$&&&$P=\frac{(z_1-z_2)^3(z_1+z_2)^3}{24\sqrt2}$\\
\ww&&&\\
$\vv2,6,3\rr=\frac{\sqrt3\h|06\ra-2\sqrt2\h|15\ra+\sqrt5\h|24\ra}{4}$&&&$P=\frac{(z_1-z_2)^5(z_1+z_2)}{16\sqrt{15}}$\\
\vspace*{-3mm}\\
\end{tabular}
\end{ruledtabular}
\end{table}

\section{discussions}

We have shown that the field-theoretical Hamiltonian describing
the interacting electrons in the LLL is invariant with respect
to the local area-preserving transformations. We have demonstrated
it for the interaction potential given by $V(r)=r^{-2\sigma}$ where
$0<\sigma<3/4$. Mathematically this is expressed by (6), which holds
for the systems with a finite electron numbers as well as in the
thermodynamic limit.

We wish to stress that the observed symmetry carries the universal
character. Namely, it persists and exhibits the unique algebraic
properties for the given class of interaction potentials. We tend
to conjecture that such an occurrence is supported by the properties
of the LLL and therefore, should be the case for a wider class of
rotationally invariant potentials. To outline these
features we have carried out an exact diagonalization explicitly in
a few body interacting electron systems. Results are summarized in
Figure 1 and Tables I and II.

Within a given block, only one eigenstate represents the ground state
while the rest ones are excitations. This may have interesting
developments involving the excitation spectra, spatial structures of
charge distribution in a ground and an excited states, etc. However,
in that case the analytic studies would be more informative than the
ones based on bare numbers. As an example, using the Table I it can
be verified that $E^{[1]}>E^{[2]}>E^{[3]}>E^{[4]}$, i.e. as if the
ground state of a given block is among the last appeared ones.
Actually, this is not always the case: $E^{[3]}>E^{[5]}>E^{[4]}$.
Therefore, the knowledge about the global tendency of eigenvalues
with respect to $\beta$ would be helpful for searching the ground
state in the thermodynamic limit. Remark, that this tendency also
seems to be insensitive to the type of $V(r)$: the above
inequalities involving $E^{[\beta]}$ hold for both types of $V(r)$
presented in Table I. In this connection one may expect the
existence of some other peculiarities which are insensitive to
a type of $V(r)$ and might shed more light on a common
analytic structure of electron-electron interactions in the LLL.

\begin{acknowledgments}

One of the authors (G.T.) expresses his sincere gratitude to
Nishina Memorial Foundation for supporting his stay at Department of
Physics, Tohoku University, where a part of this work was carried out.

M.E. and G.T. would like to thank SCOPES for the support under
grant No. 7GEPJ62379.

\end{acknowledgments}

\end{document}